\documentclass[aps,prb,twocolumn,groupedaddress,showpacs]{revtex4}

\usepackage{graphicx}
\begin{document}

\title{Landau--Zener tunneling in the presence of weak intermolecular interactions 
in a crystal of Mn$_{4}$ single-molecule magnets}

\author{W. Wernsdorfer$^1$, S. Bhaduri$^2$, A. Vinslava$^2$, and G. Christou$^2$}


\affiliation{
$^1$Lab. L. N\'eel, associ\'e \`a l'UJF, CNRS, BP 166,
38042 Grenoble Cedex 9, France\\
$^2$Dept. of Chemistry, Univ. of Florida, Gainesville, Florida 32611-7200, USA
}

\date{\today}

\begin{abstract}
A Mn$_4$ single-molecule magnet (SMM),
with a well isolated spin ground
state of $S = 9/2$, is used as a 
model system to study Landau--Zener (LZ) tunneling 
in the presence of weak intermolecular dipolar and 
exchange interactions.  The anisotropy constants $D$ and $B$ are measured with minor 
hysteresis loops. A transverse field
is used to tune the tunnel splitting over a large range.
Using the LZ and inverse LZ method, it is shown that these interactions 
play an important role in the tunnel rates.
Three regions are identified: (i) at small transverse fields,
tunneling is dominated by single tunnel transitions;
(ii) at intermediate transverse fields, the
measured tunnel rates are governed by reshuffling of internal fields,
(iii) at larger transverse fields, the magnetization reversal
starts to be influenced by the direct relaxation process,
and many-body tunnel events may occur.
The hole digging method is used to study the next-nearest neighbor 
interactions.
At small external fields, it is shown that magnetic ordering occurs which
does not quench tunneling. An applied transverse field can 
increase the ordering rate. Spin-spin cross-relaxations, mediated
by dipolar and weak exchange interactions, are proposed to
explain additional quantum steps.
\end{abstract}


\maketitle

\section{Introduction}
\label{intro} 

The nonadiabatic transition between the two states 
in a two-level system was first discussed
by Landau, Zener, and St$\ddot{\rm u}$ckelberg
~\cite{Landau32,Zener32,Stuckelberg32}. 
The original work by Zener concentrated on
the electronic states of a bi-atomic molecule, 
while Landau and St$\ddot{\rm u}$ckelberg
considered two atoms that undergo a scattering process.
Their solution of the time-dependent Schr$\ddot{\rm o}$dinger equation
of a two-level system could be applied to many physical systems,
and it became an important tool for studying tunneling transitions.
The Landau--Zener (LZ) model has also been applied to spin tunneling 
in nanoparticles and molecular clusters
~\cite{Miyashita95,Miyashita96,Rose98,Rose99,Thorwart00,Leuenberger00,Garanin02b,Garanin03a,Garanin05a}.

Single-molecule magnets~\cite{Sessoli93b,Sessoli93,Aubin96} 
have been the most promising spin systems to date for 
observing quantum phenomena like Landau--Zener 
tunneling because they have a 
well-defined structure with well-characterized 
spin ground state and magnetic anisotropy~\cite{WW_Science99,WW_PRL05}. 
These molecules can be assembled in 
ordered arrays where all molecules have 
the same orientation. Hence, macroscopic 
measurements can give direct access to 
single molecule properties. 

Since SMMs occur as assemblies in crystals, there is the 
possibility of a small electronic interaction
of adjacent molecules. This leads to very 
small exchange interactions that depend strongly on the 
distance and the non-magnetic atoms 
in the exchange pathway. 
Up to now, such an intermolecular exchange interaction 
has been assumed to be negligibly small. However, our 
recent studies on several SMMs suggest 
that in most SMMs exchange interactions 
lead to a significant influence on the tunnel process~\cite{WW_PRL02}. 
Recently, this intermolecular exchange 
interaction was used to couple antiferromagnetically 
two SMMs, each acting as a bias on its 
neighbor~\cite{WW_Nature02,TironPRB03,TironPRL03,Hill_Science03,Yang03}. 

In this paper we present a detailed study of Landau-Zener
tunneling in a Mn$_4$ SMM with a well isolated spin ground
state of $S = 9/2$. Using the standard and 
the inverse LZ method we show that spin-spin interactions
are strong in SMMs with large tunnel splittings.
By applying transverse fields, we can tune the tunnel
splittings from kHz to sub-GHz tunnel frequencies. 
We identify three regions depending
on the applied transverse field.
Next-nearest neighbor interactions, ordering,
and spin-spin cross relaxations are studied.

Several reasons led us to the choice of this SMM:
(i) a half integer spin is very convenient
to study different regions of tunnel splittings.
At zero applied field, the Kramers degeneracy is
only lifted by internal fields (dipolar, exchange,
and nuclear spin interactions). A transverse field
can then be used to tune the tunnel splitting over
a large range;
(ii) Mn$_4$ has a spin ground state S = 9/2 well separated from the
first excited multiplet (S=7/2) by about 300 K~\cite{Aubin96,Aubin98b};
(iii) Mn$_4$ has one of the largest uniaxial anisotropy 
constants, $D$, leading to well separated tunnel resonances;  
(iv) the spin ground state $S$ is rather small allowing easy
studies of ground state tunneling; 
(v) Mn$_4$ has a convenient crystal symmetry leading to needle 
shaped crystals with the easy axis of magnetization
(being the $c$-axis) along the 
longest crystal direction; and (vi)
the weak spin chain-like exchange and dipolar interactions 
of Mn$_4$ are well controlled.

\begin{figure*}
\includegraphics[width=.65\textwidth]{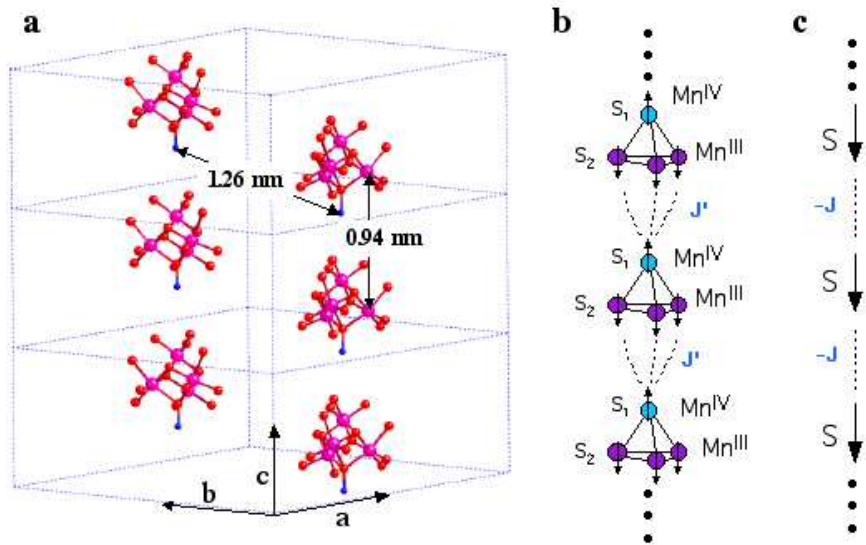}
\caption{(color online) (a) Unit cells of the Mn$_4$ crystal. 
Only the cores of the Mn$_4$ molecules are shown. The largest
spheres are Mn, the smallest Si, and the others O atoms.
The distances between next-nearest molecules are indicated.
(b) Schematic drawing of the chain-like coupling between the Mn4 SMMs.
(c) Scheme of the chain model where the $S = 9/2$ spin
of each molecule is represented by an arrow.}
\label{structure_cell_SB1}
\end{figure*}

\section{Structure of ${\rm Mn}_4$ and measuring technique}
\label{structure} 

The studied SMM has the formula 
[Mn$_4$O$_3$(OSiMe$_3$)(OAc)$_3$(dbm)$_3$], 
called briefly Mn$_4$. The preparation, 
X-ray structure, and detailed physical characterization 
are reported elsewhere~\cite{Bhaduri03}. 
Mn$_4$ crystallizes in a hexagonal space group 
with crystallographic C$_{\rm 3}$ symmetry.
The unit cell parameters are
$a$ = $b$ = 1.998 nm, $c$ = 0.994 nm, 
$\alpha = \beta = 90^{\circ}$, and $\gamma = 120^{\circ}$.
The unit cell volume is 3.438 nm$^3$ and two molecules
are in a unit cell.
The complex has a distorted cubane-like core geometry and 
is Mn$_3^{\rm III}$Mn$^{\rm IV}$. 
The C$_{\rm 3}$ axis passes through the Mn$^{\rm IV}$ ion 
and the triply bridging siloxide group (Fig.~\ref{structure_cell_SB1}a). 
DC and AC magnetic susceptibility measurements 
indicate a well isolated $S = 9/2$ ground state~\cite{Bhaduri03}. 

We found a fine structure of three in
the zero-field resonance (Sect.~\ref{fine_structure_of_three}) that is due to 
the strongest nearest neighbor 
interactions of about 0.036 T along the $c-$axis 
of the crystals. This coincides with the shortest Mn--Mn
separations of 0.803 nm between two molecules along the $c-$axis, 
while the shortest Mn--Mn separations 
perpendicular to the $c-$axis are 1.69 nm
and in diagonal direction 1.08 nm (Fig.~\ref{structure_cell_SB1}a).
We cannot explain the value of 
0.036 T by taking into account only dipolar 
interactions, which should not be larger 
than about 0.01 T. We believe therefore that small 
exchange interactions are responsible for the observed value.
Indeed, the SMMs are held together by three H bonds C--H--O
which are probably responsible for the small exchange interactions.

Fig.~\ref{structure_cell_SB1}b shows schematically the antiferromagnetic
exchange coupling between the Mn$^{\rm IV}$ $(S_1=3/2)$ ions of one molecule 
and the Mn$^{\rm III}$ $(S_2=2)$ ion of the neighboring molecule, going via 
three H bonds C--H--O (not shown in Fig.~\ref{structure_cell_SB1}b). 
This leads to an effective ferromagnetic coupling between the 
collective spins $(S = 9/2)$ of the SMMs (Fig.~\ref{structure_cell_SB1}c) 
because the  Mn$^{\rm IV}$ $(S_1=3/2)$ ions 
and the Mn$^{\rm III}$ $(S_2=2)$ ion in each 
molecule are antiferromagnetically coupled.

All measurements were performed using an 
array of micro-SQUIDs~\cite{WW_ACP01}. 
The high sensitivity  allows us to study single 
crystals of SMMs of the order of 5 $\mu$m or larger. 
The field can be applied in any direction by separately 
driving three orthogonal coils.
In the present study, the field was always aligned 
with the C$_{\rm 3}$ axis of the molecule, 
that is the magnetic easy axis, 
with a precision better than 0.1$^{\circ}$~\cite{WW_PRB04}.
The transverse fields were applied transverse to the 
C$_{\rm 3}$ axis and along the $a-$axis.

\section{Spin Hamiltonian and Landau--Zener tunneling}
\label{LZ} 

\begin{figure}
\begin{center}
\includegraphics[width=.45\textwidth]{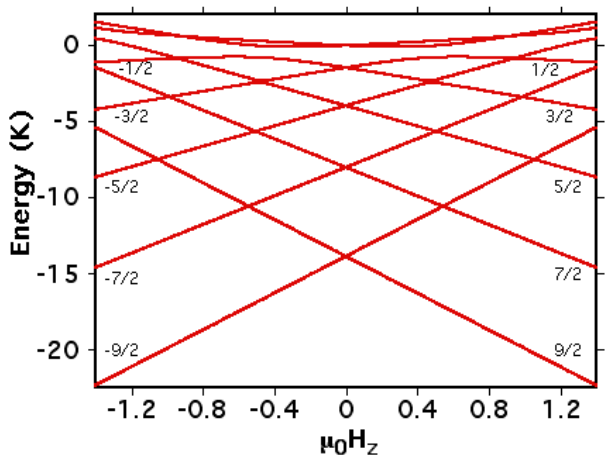}
\caption{(color online)  Zeeman diagram of the 10 levels of 
the $S = 9/2$ manifold of Mn$_4$ as a 
function of the field applied along the easy axis. 
The spin Hamiltonian parameters are $D$ = 0.608 K,
$B$ = 3.8 mK, and $E$ = 0.032 K.
The levels are approximately labeled 
with quantum numbers $m = \pm9/2, \pm7/2, ..., \pm1/2$. }
\label{Zeeman}
\end{center}
\end{figure}

The single spin model (giant spin model) 
is the simplest model describing the spin 
system of an isolated SMM. The spin Hamiltonian is
\begin{equation}
	\mathcal{H} = -D S_{z}^2 -B S_{z}^4 + \mathcal{H}_{{\rm trans}} 
	- g \mu_{\rm B} \mu_0 \vec{S}\cdot\vec{H} 
\label{eq_H}
\end{equation}
$S_{x}$, $S_{y}$, and $S_{z}$ are the
components of the spin operator, $g~\approx~2$, $\mu_{\rm B}$ the Bohr magneton; 
$D$ and $B$ the anisotropy constant defining an 
Ising type of anisotropy; $\mathcal{H}_{{\rm trans}}$, 
containing $S_{x}$ or $S_{y}$ spin operators, 
gives the transverse anisotropy which is small 
compared to $D S_{z}^2$ in SMMs; and the last term 
describes the Zeeman energy associated 
with an applied field $\vec{H}$. 
This Hamiltonian has an energy level spectrum 
with $(2S+1)$ values which, to a first approximation, 
can be labeled by the quantum numbers $m = -S, -(S-1), ...,  S$ 
taking the $z$-axis as the quantization axis. 
The energy spectrum can be obtained by using standard 
diagonalization techniques (Fig.~\ref{Zeeman}). 
At $\vec{H} = 0$, the levels $m = \pm S$ have the 
lowest energy. When a field $H_z$ is applied, 
the levels with $m > 0$ decrease in energy, 
while those with $m < 0$ increase. Therefore, 
energy levels of positive and negative quantum numbers 
cross at certain values of $H_z$.
Although $\mathcal{H}_{\rm trans}$ produces
tunneling, it can be neglected when determining the field
positions of the level crossing because it is often much smaller than the
axial terms. Without $\mathcal{H}_{\rm trans}$
and transverse fields, the
Hamiltonian is diagonal and 
the field position of the crossing of level 
$m$ with $m'$ is given by
\begin{equation}
	H_{m,m'}=\frac {n\left[D+B\left(m^2+m'^2\right)\right]}
	{g\mu_{\rm B}\mu_0 } 
\label{eq_H_step}
\end{equation}
where $n = -(m+m')$ is the step index.

When the spin Hamiltonian contains transverse terms 
($\mathcal{H}_{\rm trans}$), the level crossings  can 
be {\it avoided level crossings}. The spin $S$ is {\it in resonance} 
between two states when the local longitudinal 
field is close to an avoided level crossing. 
The energy gap, the so-called {\it tunnel splitting}
$\Delta$, can be tuned by a transverse field 
(perpendicular to the $S_z$ 
direction)~\cite{Garg93,WW_Science99,WW_PRL05}.

The nonadiabatic tunneling probability $P_{m,m'}$ between two states 
when sweeping the longitudinal field $H_z$ at a constant rate 
over an avoided energy level crossing was first discussed
by Landau, Zener, and St$\ddot{\rm u}$ckelberg
~\cite{Landau32,Zener32,Stuckelberg32}.
It is given by
\begin{equation}
P_{m,m'} = 1 - {\rm exp}\left\lbrack-
     \frac {\pi \Delta_{m,m'}^2}
     {2 \hbar g \mu_{\rm B} |m - m'| \mu_0 dH_z/dt}\right\rbrack
\label{eq_LZ}
\end{equation}
Here, $m$ and $m'$ are the quantum numbers of the avoided level crossing, 
$dH_z/dt$ is the constant field sweeping rates, 
and $\hbar$ is Planck's constant. 

\begin{figure}
\includegraphics[width=.4\textwidth]{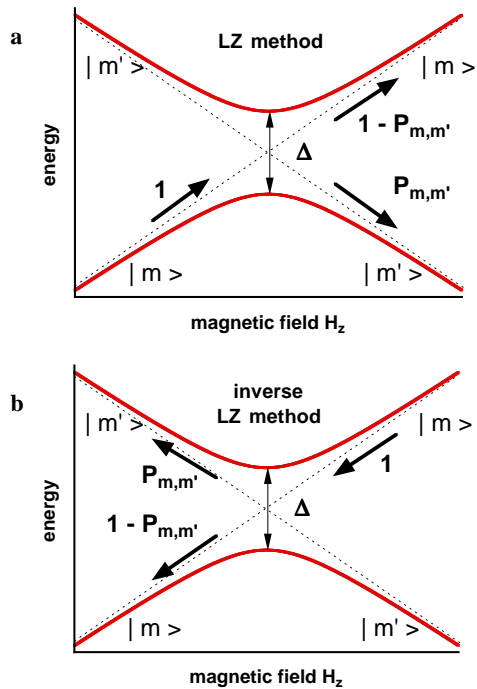}
\caption{(color online) Detail at a level crossing $m$ 
with $m'$ where the 
transverse terms (terms containing $S_x$ 
or/and $S_y$ spin operators) turn the crossing 
into an avoided level crossing. 
The initial state is the lower
energy state in (a) (standard LZ method) whereas in (b) 
it is the higher energy state (inverse LZ method).}
\label{LZ_crossings}
\end{figure}

Fig.~\ref{LZ_crossings} presents two different methods 
to apply the LZ model: in Fig.~\ref{LZ_crossings}a, 
the initial state is the lower
energy state (standard LZ method) whereas in Fig.~\ref{LZ_crossings}b 
it is the higher energy state (inverse LZ method). 
The tunneling probabilities are given by Eq.~\ref{eq_LZ}. 
In the simple LZ scheme, both methods should lead to the same 
result. However, when introducing interactions of the
spin system with environmental degrees of freedom (phonons,
dipolar and exchange interactions, nuclear spins, etc.),
both methods are quite different because the final state in 
Fig.~\ref{LZ_crossings}a
and the initial state in Fig.~\ref{LZ_crossings}b are unstable. 
The lifetimes of these
states depend on the environmental couplings as well as the level
mixing which can be tuned with an applied transverse field.
We will see in Sect.~\ref{LZ_Mn4} that comparison of
the two methods allows the effect of environmental interactions
to be observed.

\begin{figure}
\includegraphics[width=.45\textwidth]{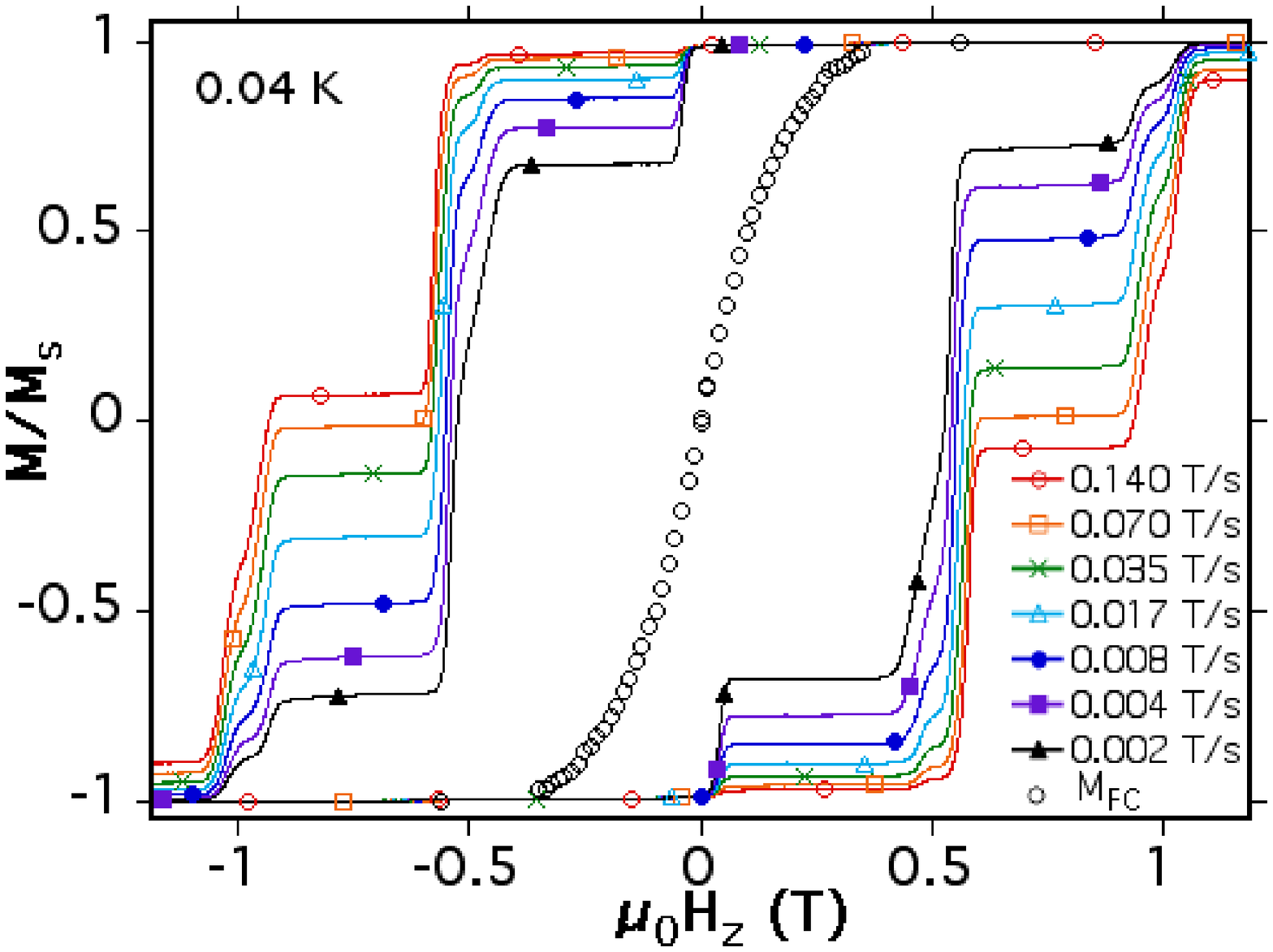}
\includegraphics[width=.45\textwidth]{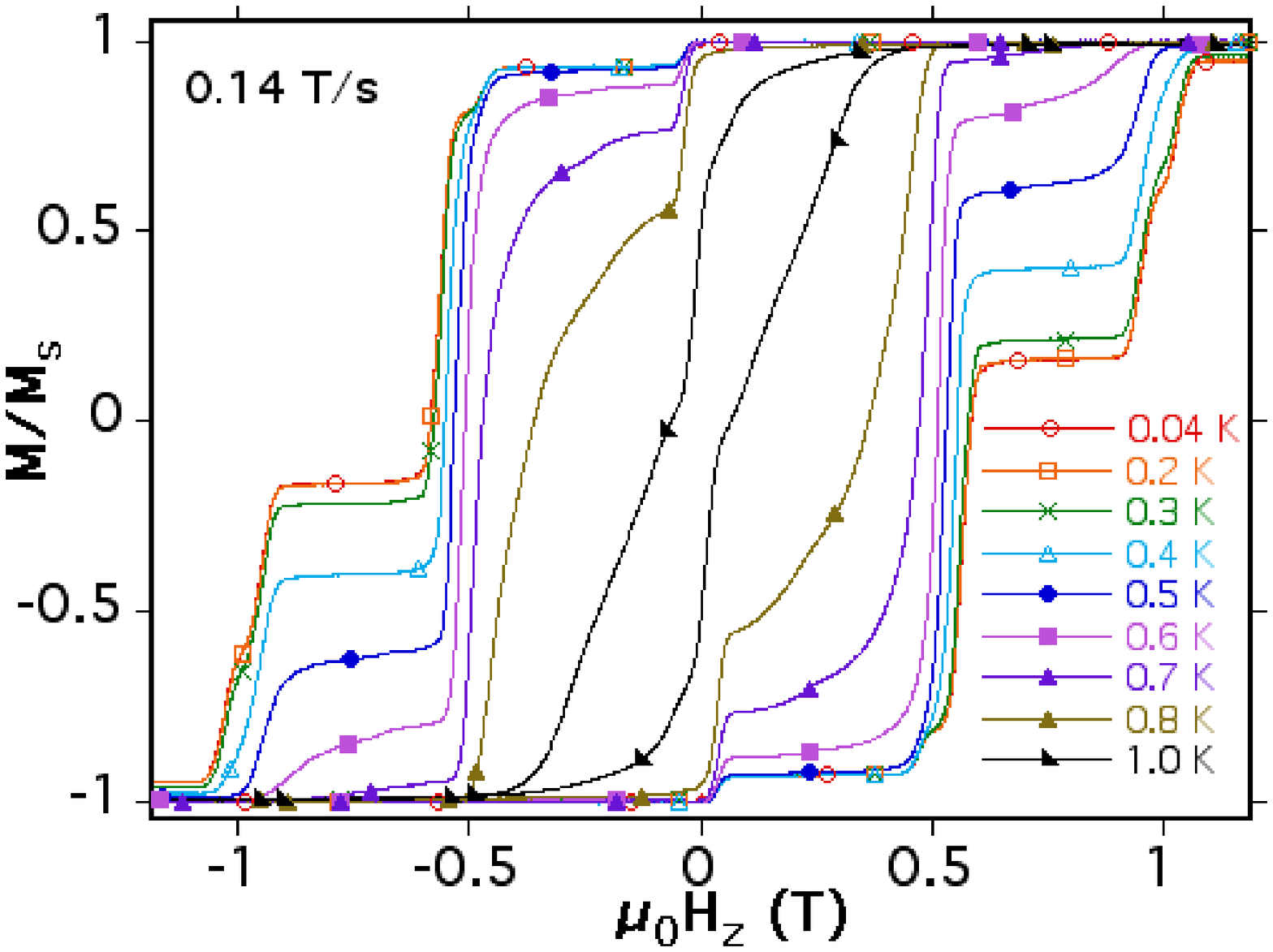}
\caption{(color online) (a) Hysteresis loop measurements of a single crystal of
Mn$_4$ at low temperatures (40 mK) where thermal activation to
excited spin states can be neglected. The field is applied in the
direction of the easy axis of magnetization and swept at a
constant rate between 0.002 and 0.14 T/s. 
The dots labeled with $M_{\rm FC}$ are the magnetization after
cooling the sample from 5 K down to 0.04 K in a constant applied field 
$H_z$. $M_{\rm FC}$ is used for the equilibrium magnetization
$M_{\rm eq}$ in Sect.~\ref{digging}.
(b) Hysteresis loop measurements similar to Fig. 2 but at
different temperatures and for a field sweep rate of 0.14 T/s.}
\label{hyst}
\end{figure}

\subsection{Landau--Zener tunneling in Mn$_4$}
\label{LZ_Mn4} 

Landau--Zener tunneling can 
be seen in hysteresis loop measurements. 
Figs.~\ref{hyst}a and~\ref{hyst}b show typical hysteresis loops 
for a single crystals of Mn$_4$ at several temperatures
and field sweep rates. 
When the applied field is near an 
avoided level crossing, the magnetization relaxes faster, 
yielding steps separated by plateaus. 
As the temperature is lowered, there is a decrease 
in the transition rate due to reduced thermally assisted tunneling. 
A similar behavior was observed in Mn$_{12}$ acetate clusters 
\cite{Novak95,Paulsen95,Paulsen95b,Friedman96,Thomas96} 
and other 
SMMs~\cite{Sangregorio97,Aubin98,Caneschi99,Yoo_Jae00,Ishikawa05b}. 
The hysteresis loops become temperature-independent below 0.4 K
indicating ground state tunneling. The field between two resonances
allows us to estimate the anisotropy constants $D$ and $B$. 
We found:
\begin{equation}
D = g\mu_{\rm B}\mu_0\left(H_z^{(1)}- \frac{2S^2-2S+1}{2S-3}
      \left(H_z^{(1)}-\frac{ H_z^{(2)}}{2}\right)\right)
\label{eq_D}
\end{equation}
\begin{equation}
B = \frac{g\mu_{\rm B}\mu_0}{2S-3}
       \left(H_z^{(1)}-\frac{H_z^{(2)}}{2}\right)
\label{eq_B}
\end{equation}
where $H_z^{(1)}$ and $H_z^{(2)}$ are the field positions 
of level crossings $M = -S$ with $S-1$ 
and $M = -S$ with $S-2$. 

\begin{figure}
\begin{center}
\includegraphics[width=.45\textwidth]{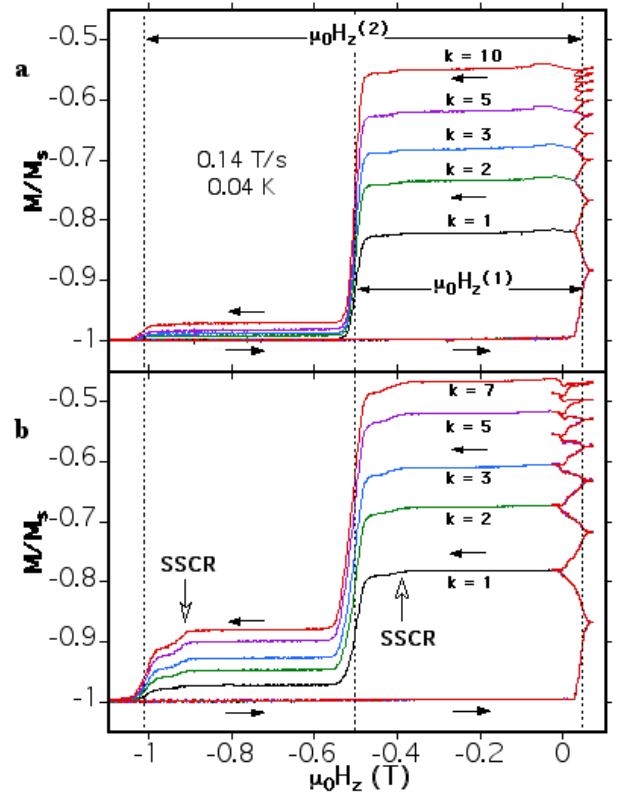}
\caption{(color online)  Minor hysteresis loops of a single crystal of Mn$_4$.
The magnetization was first saturated at -1.4 T. 
After ramping the field to zero at 0.14 T/s, the field
was swept $k$-times back and forth 
(between 0.028 and 0.07 T in (a) and between
-0.028 and 0.07 T in (b))
over the zero-field resonance with a sweep rate of 0.014 T/s.
After the $k$-th sweep, the field is quickly
swept back to -1.4 T at a rate of 0.14 T/s
leading to resonant tunneling at the 
level crossing $(m,m')=(-7/2,9/2)$ and $(-5/2,9/2)$,
and allow to determine $\mu_0H_z^{(1)}$ = 0.544 T 
and $\mu_0H_z^{(2)}$ = 1.054 T. 
The field interval of the $k$ back and forth sweeps 
corresponds to {\it zero reversed neighbor} (0 RN, see Sect.~\ref{fine_structure_of_three})
in (a) whereas it goes over the 0 RN, 1 RN, and 2 RN
transitions in (b).
Note that the procedure in (a) leads to sharper steps and
reduce spin-spin cross-relaxtion (SSCR) 
(Sect.~\ref{SSCR}) because all reversed
spins have two non-reversed neighbors.
The transitions of SSCR are indicated in (b).}
\label{hyst_min}
\end{center}
\end{figure}

The influence of dipolar and intermolecular exchange, 
which can shift slightly 
the resonance positions (Sect.~\ref{fine_structure_of_three}), 
can be avoided by performing minor hysteresis loops involving only 
few percent of the molecules (Fig.~\ref{hyst_min}). We found the field 
separations between the zero-field resonance 
and the first and second resonance are 
$H_z^{(1)}$ = 0.544 T and $H_z^{(2)}$ = 1.054 T. 
Using Egs.~\ref{eq_D} and~\ref{eq_B}, we find
$D$ = 0.608 and $B$ = 3.8 mK.
These values agree with those obtained from 
INS and EPR measurements~\cite{Hill_unpublished}.

In order to explain the few minor steps (Fig.~\ref{hyst}), not 
explained with the above Hamiltonion, 
spin-spin cross-relaxation between adjacent 
molecules has to be taken into account~\cite{WW_PRL02}.
Such relaxation processes, 
present in most SMMs, are well resolved for Mn$_4$
because the spin is small (Sect.~\ref{SSCR}).

The spin-parity effect was established by measuring
the tunnel splitting $\Delta$ as a function of transverse 
field because $\Delta$ is expected
to be very sensitive to
the spin-parity and the parity of the avoided level crossing. 
We showed elsewhere that the tunnel splitting increases gradually for an integer spin,
whereas it increases rapidly for a half-integer spin~\cite{WW_PRB02}.
In order to apply quantitatively the LZ formula 
(Eq.~\ref{eq_LZ}), 
we first checked the predicted 
field sweep rate dependence of the tunneling rate. 
The SMM crystal was placed in a high 
negative field to saturate the magnetization, 
the applied field was swept at a constant rate 
over one of the resonance transitions, and 
the fraction of molecules that 
reversed their spin was measured. The tunnel splitting $\Delta$ 
was calculated using Eq.~\ref{eq_LZ} and was plotted 
in Fig. 2 of reference $\cite{WW_PRB02}$ as a function of field sweep rate. The LZ method
is applicable in the region of high sweep rates 
where $\Delta_{-9/2,9/2}$ is independent of the field sweep rate.
The deviations 
at lower sweeping rates are mainly due to 
reshuffling of internal fields~\cite{WW_PRL99} (Sect.~\ref{digging}) 
as observed for the Fe$_8$ SMM~\cite{WW_EPL00}.
Such a behavior has recently been simulated~\cite{Liu01}.
\footnote{A more detailed study shows that the tunnel
splittings obtained by the LZ method
are slightly influenced by environmental effects
like hyperfine and dipolar couplings~\cite{WW_EPL00,Liu01}. 
Therefore, one might 
call it an effective tunnel splitting.
}

\subsubsection{LZ tunneling in the limit of $P_{m,m'} << 1$}
\label{LZ_Mn4_P_small} 

Fig. 3 of reference $\cite{WW_PRB02}$ presents the tunnel splittings obtained by 
the LZ method as a function of transverse
field and shows that the tunnel splitting increases 
rapidly for a half-integer spin.
Figs. 4a and 4b of reference $\cite{WW_PRB02}$ present a simulation of the 
measured tunnel splittings. 
We found that either
the second order term ($E(S_+^2+S_-^2)$) 
with $E$ = 0.032 K or a fourth
order term ($B_{44}(S_+^4+S_-^4)$) 
with $B_{44}$ = 0.03 mK can equally  well
describe the experimental data.
These results suggest that  there is a small effect that breaks
the  C$_3$ symmetry. This could  be a small strain inside the
SMM crystal induced by defects, which could result
from a loss or disorder of solvent molecules. 
Recent inelastic neutron scattering
measurements confirm the presence of second and 
fourth order terms~\cite{Bhaduri03}.

\begin{figure}
\includegraphics[width=.45\textwidth]{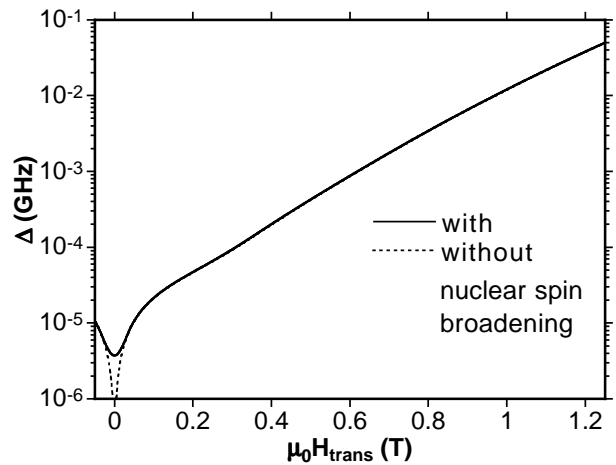}
\caption{(color online)  Calculated tunnel splitting for Mn$_4$ 
using $D$ = 0.608 K, $B$ = 3.8 mK, and E = 0.038 K $\cite{WW_PRB02}$.
The calculated $\Delta$ has been 
averaged over all possible orientations 
of the transverse field in order to represent
the arbitrary orientation of the $E$ term.
The influence of nuclear spin broadening
was taken into account by a Gaussian 
distribution of transverse field components with a 
half-width $\sigma =$ 0.035 T.}
\label{delta_GHz}
\end{figure}

\begin{figure}
\includegraphics[width=.45\textwidth]{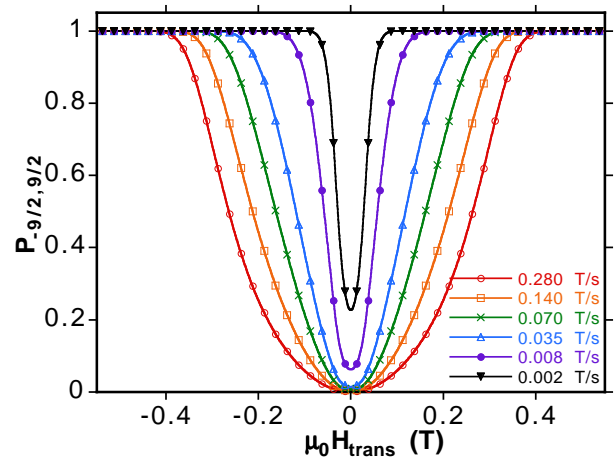}
\caption{(color online)  Calculated 
LZ tunnel probabilities $P_{\pm9/2}$
as a function of transverse field using the 
tunnel splitting from Fig.~\ref{delta_GHz} and 
the indicated field sweep rates. 
Only every 100-th calculated point is shown as a symbol.}
\label{P_LZ}
\end{figure}

\begin{figure*}
\includegraphics[width=.85\textwidth]{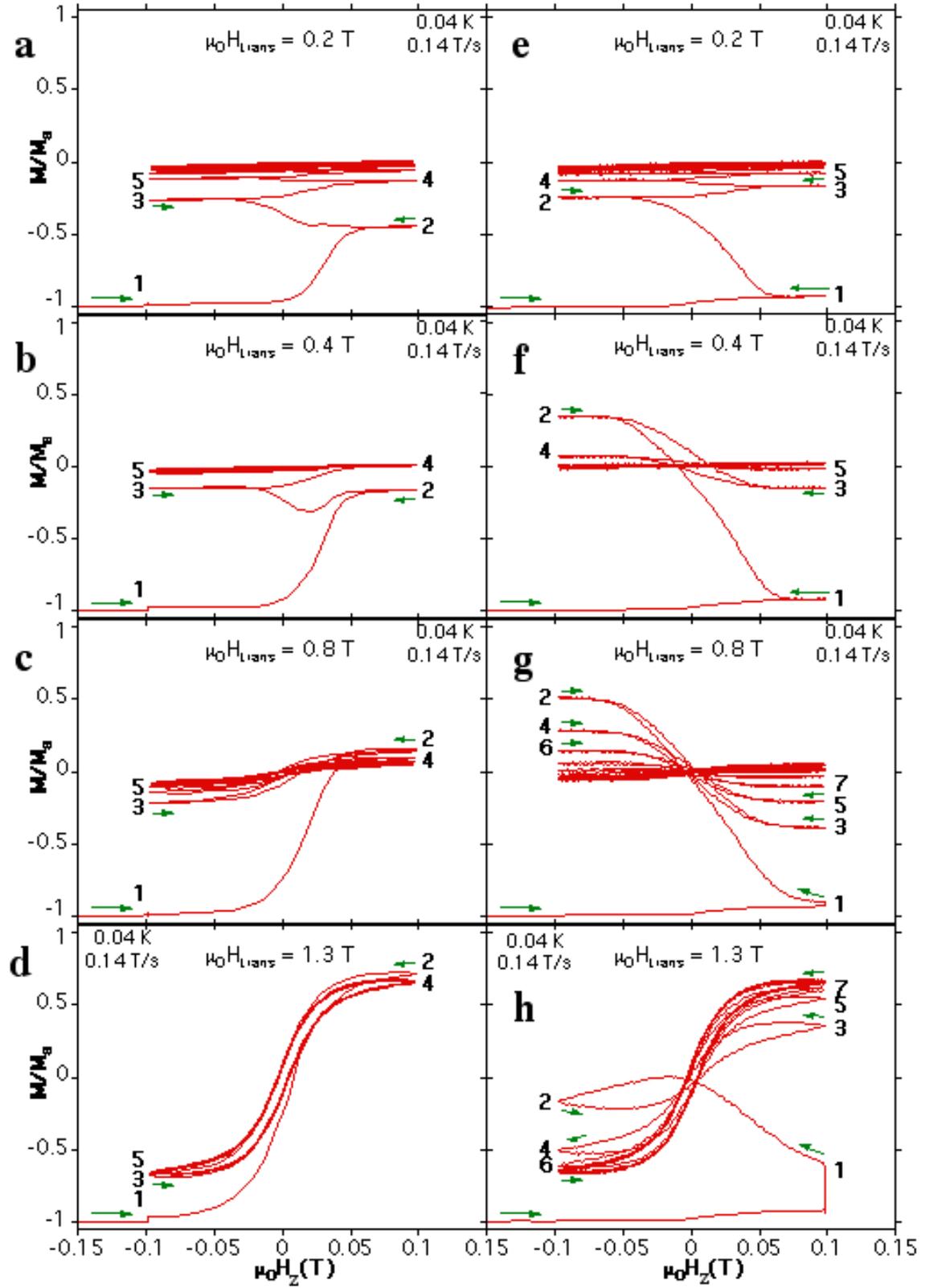}
\caption{(color online)  Magnetization versus applied field scans
for (a-d) the LZ method and (e-h) the inverse LZ method.
The indicated transverse fields $H_{\rm trans}$ are
applied at point {\bf 1}.}
\label{cy_LZ}
\end{figure*}

\subsubsection{LZ tunneling  for large probabilities $P_{m,m'}$}
\label{LZ_Mn4_P_large} 
Figs.~\ref{delta_GHz} and~\ref{P_LZ} show respectively 
the tunnel splitting $\Delta$ and the LZ tunnel probabilities $P_{\pm9/2}$
as a function of transverse field using the parameters 
of reference $\cite{WW_PRB02}$ 
(Sect.~\ref{LZ_Mn4_P_small}). $P_{\pm9/2}$ increases rapidly
to unity; for example, $P_{\pm9/2}$ = 1 for $H_{\rm trans}>0.4$~T
and field sweep rates $dH/dt$ smaller than 0.28 T/s.
The Mn$_4$ system is therefore ideal to study different
regimes of the tunnel probability ranging from 
kHz to sub-GHz tunnel frequencies (Figs.~\ref{delta_GHz}).

Fig.~\ref{cy_LZ} presents the magnetization 
variation during LZ field sweeps 
for several transverse fields.
The SMM crystal was first placed in a high 
negative field to saturate the magnetization and 
the applied field was then swept at a constant rate 
to the field value of -0.1 or 0.1~T for 
the LZ (Fig.~\ref{cy_LZ}a-d) or inverse LZ
method (Fig.~\ref{cy_LZ}e-h). At this field
value, labeled {\bf 1}, a transverse field was
applied to increase the tunnel probability. 
Finally, the field is swept back
and forth over the zero-field resonance transitions
($m = \pm9/2$) and  the fraction of molecules 
that reversed their spin was measured. 

Note severals points in Fig.~\ref{cy_LZ}:
(i) in Figs.~\ref{cy_LZ}a and~\ref{cy_LZ}e, the magnetization increases
gradually with each field sweep from $M = -M_{\rm s}$ to $M = 0$;
(ii) in Fig.~\ref{cy_LZ}b, the field sweep from {\bf 2} to {\bf 3}
shows first a decrease and then an increase of magnetization.
This is due to next-nearest neighbor effects (Sect.~\ref{fine_structure_of_three});
(iii) in Figs.~\ref{cy_LZ}c and~\ref{cy_LZ}d, the magnetization increases
(decreases) for a positive (negative) field scan. Note that
the tunnel probability is 1 for transverse fields larger
than 0.3 T (Figs.~\ref{P_LZ}), that is all spins should reverse
for each field sweep;
(iv) in Figs.~\ref{cy_LZ}f and~\ref{cy_LZ}g, the magnetization increases
much stronger for the field sweep from {\bf 1} to {\bf 2} than
in Figs.~\ref{cy_LZ}b and~\ref{cy_LZ}c;
(v) in Fig.~\ref{cy_LZ}h, the field sweep from {\bf 1} to {\bf 2}
shows first an increase and then a decrease of magnetization;
(vi) in Figs.~\ref{cy_LZ}a to~\ref{cy_LZ}c and~\ref{cy_LZ}e to~\ref{cy_LZ}g,
the magnetization tends to relax towards $M$ = 0 whereas
in Figs.~\ref{cy_LZ}d and~\ref{cy_LZ}h, it relaxes towards
the field cooled magnetization $M_{\rm FC}$ (Fig.~\ref{hyst}a).

\begin{figure}
\includegraphics[width=.45\textwidth]{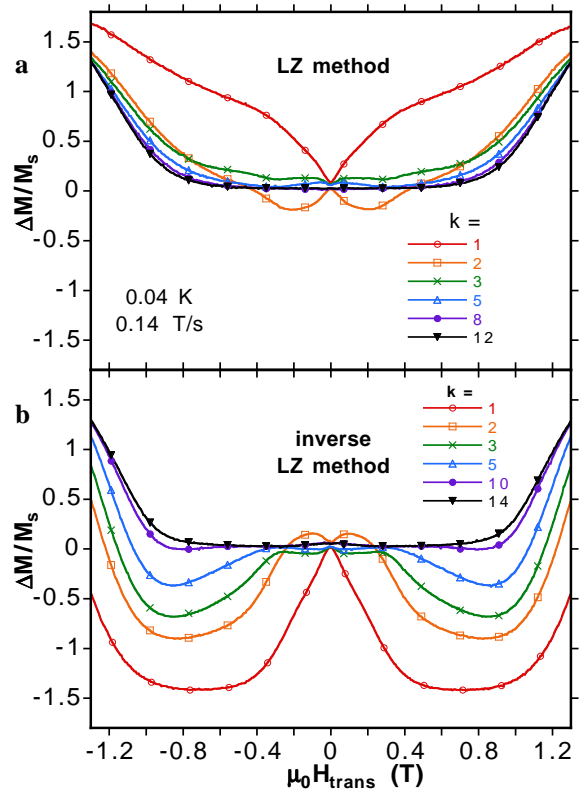}
\caption{(color online)  The change of magnetization 
$\Delta M = (M_{\rm f}-M_{\rm i})\frac{dH}{dt}/|\frac{dH}{dt}|$ 
as a function of transverse field $H_{\rm trans}$ for several
LZ field sweeps; (a) LZ method and (b) inverse LZ method.
Only every 30-th measured point is shown as a symbol.}
\label{deltaM_H_tr_k}
\end{figure}

\begin{figure}
\includegraphics[width=.45\textwidth]{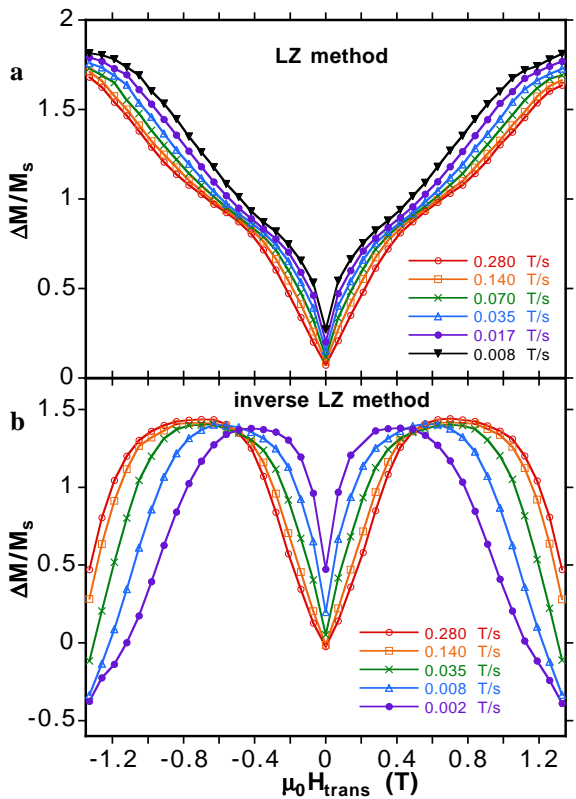}
\caption{(color online)  The change of magnetization 
$\Delta M = (M_{\rm f}-M_{\rm i})\frac{dH}{dt}/|\frac{dH}{dt}|$ 
as a function of transverse field $H_{\rm trans}$ for
the LZ field sweep from {\bf 1} to {\bf 2} and for the indicated
field sweep rates; (a) LZ method and (b) inverse LZ method.}
\label{deltaM_H_tr_v}
\end{figure}

The result of a detailed study of the magnetization
change $\Delta M$ for LZ field scans like those in Fig.~\ref{cy_LZ}
are summarized in Figs.~\ref{deltaM_H_tr_k} and~\ref{deltaM_H_tr_v}.
$\Delta M$ is obtained from
$\Delta M = (M_{\rm f}-M_{\rm i})\frac{dH}{dt}/|\frac{dH}{dt}|$
where $M_{\rm i}$ and $M_{\rm f}$ are the initial and final
magnetization for a given LZ field sweep.
Fig.~\ref{deltaM_H_tr_v} gives field sweep rate dependence
for the field sweep from {\bf 1} to {\bf 2}.
These graphs show clearly the crossover between the 
different regions presented in Fig.~\ref{cy_LZ}.

We identify three regions: 

(i) at small transverse fields
(0 to 0.2 T), that is $P_{\pm9/2} << 1$,
tunneling is dominated by single tunnel transitions
and $\Delta M$ follows the LZ formula (Eq.~\ref{eq_LZ}).
This regime is described in Sect.~\ref{LZ_Mn4_P_small};

(ii) at intermediate transverse fields (0.2 to 0.7 T),
that is tunnel probabilities $P_{\pm9/2}$ between $\approx$0.1
and $\approx$1, $\Delta M$ deviates strongly from Eq.~\ref{eq_LZ}
and is governed by reshuffling of internal fields;

(iii) at larger transverse fields, the magnetization reversal
starts to be influenced by the direct relaxation process~\cite{Abragam70}
and many-body tunnel events may occur.

The dominating reshuffling of internal fields in region (ii) can be
seen when one compares $\Delta M$
in Figs.~\ref{cy_LZ}b and~\ref{cy_LZ}c 
for the field sweep from {\bf 1} to {\bf 2} with those
in Figs.~\ref{cy_LZ}f and~\ref{cy_LZ}g.
A backward sweep gives a larger step than a forward sweep.
This is expected for a weak ferromagnetically coupled spin chain.
Indeed, any spin that reverses shifts (shuffles) its neighboring spins
to negative fields. For a forward sweep this means that these
spins will not come to resonance whereas in a backwards sweep,
these spins might tunnel a little bit late during the field sweep.
A more detailed discussion is presented in Sect.~\ref{fine_structure_of_three}

In region (iii) the direct relaxation process~\cite{Abragam70} between the two
lowest levels starts to  play a role. 
This can be seen by the fact that,
during the application of the transverse field in point {\bf 1}
(Figs.~\ref{cy_LZ}h), the magnetization starts to relax rapidly.
A direct relaxation process is indeed probable when the 
involved levels start to be mixed by the large
transverse field. Because of this level mixing and the 
intermolecular interactions, multi-tunnel events are possible
because neighboring spins start to be entangled. 

The inverse LZ method allows us to establish adiabatic
LZ transitions. Whereas for the standard LZ method 
the difference between an adiabatic and strongly decoherent
transition is difficult to distinguish, the inverse
LZ method allows a clear separation. This is due to the
fact that the equilibrium curve and the adiabatic 
curve are similar for the standard LZ method but
not for the inverse LZ method. For example, Fig.~\ref{cy_LZ}g
shows that there are more than 10 adiabatic LZ passages
before the system reaches a disordered state.
It is difficult to conclude this from Fig.~\ref{cy_LZ}c
because strong decoherence would lead to a similar curve.

It is important to note that the transition between 
regions (ii) and (iii) leads to the shoulder in Fig.~\ref{deltaM_H_tr_v}
which should not be interpreted as quantum phase interference. 

Fig.~\ref{deltaM_H_tr_k} and Fig.~\ref{deltaM_H_tr_v}
are very rich with information and a complete understanding
needs a multi-spin simulation.

\section{Intermolecular dipolar and exchange interactions}
\label{interaction} 
SMMs can be arranged in a crystal with all molecules
having the same orientation. Typical distances between molecules 
are between 1 and 2 nm. Therefore, intermolecular dipole interactions
cannot be neglected. An estimation of the dipolar energy can be 
found in the  mean field  approximation.
\begin{equation}
E_{\rm dip} = \frac{\mu_0}{4 \pi}
                        \frac {\left(g \mu_{\rm B} S \right)^2}{V}
\label{E_dip}
\end{equation}
where $V$ is the volume of 
the unit cell divided by the number of molecules per unit cell.
Typical  values of $E_{\rm dip}$ for SMM are between
0.03 and 0.2 K.
More precise values, between 0.1 and 0.5 K, 
were calculated recently~\cite{Chudnovsky01b,Fernandez02}.

In addition to dipolar interactions there is also the 
possibility of a small electronic interaction
of adjacent molecules. This leads to very 
small exchange interactions that depend strongly on the 
distance and the non-magnetic atoms 
in the exchange pathway. 
Until recently, such intermolecular exchange interactions 
have been assumed to be negligibly small. However, our 
recent studies on several SMMs suggest 
that in most SMMs exchange interactions 
lead to a significant influence on the tunnel 
process~\cite{WW_Nature02,TironPRB03,TironPRL03,Hill_Science03,Yang03}. 

The main difference between dipolar and exchange 
interactions are: (i) dipolar interactions are long 
range whereas exchange interactions are usually short range; 
(ii) exchange interactions can be much stronger 
than dipolar interactions; (iii) whereas the sign 
of a dipolar interaction can be determined easily, 
that of exchange depends strongly on electronic 
details and is very difficult to predict; and
(iv) dipolar interactions depend strongly on the 
spin ground state $S$, whereas exchange interactions 
depend strongly on the single-ion spin states. 
For example, intermolecular dipolar interactions 
can be neglected for antiferromagnetic SMMs 
with $S = 0$, whereas intermolecular exchange 
interactions can still be important and 
act as a source of decoherence.

\subsection{Hole digging method to study 
intermolecular interactions}
\label{digging} 

Here, we focus on the 
low temperature and low field limits, 
where phonon-mediated relaxation is astronomically long and can be 
neglected. In this limit, the $m~= \pm S$ spin 
states are coupled due to the tunnel 
splitting $\Delta_{\pm S}$ which is about 
10$^{-7}$~K for Mn$_4$ (Sect.~\ref{LZ}). 
In order to tunnel between these 
states, the longitudinal magnetic energy 
bias $\xi=g\mu_{\rm B}SH_{\rm local}$ due to the local 
magnetic field $H_{\rm local}$ on a molecule must be smaller than $\Delta_{\pm S}$,
implying a local field smaller than $10^{-7}$~T for Mn$_4$ 
clusters. Since the typical intermolecular dipole fields for Mn$_4$ are of the 
order of 0.01~T and the exchange field between two adjacent
molecules of the order of 0.03~T, 
it seems at first that almost all molecules should be 
blocked from tunneling by a very large energy bias. Prokof'ev and Stamp 
have proposed a solution to this dilemma by proposing that fast dynamic 
nuclear fluctuations broaden the resonance, and the gradual adjustment 
of the internal fields in the sample caused by the tunneling brings 
other molecules into resonance and allows continuous relaxation 
\cite{Prokofev98}. 

Prokof'ev and Stamp showed that at a 
given longitudinal applied field $H_z$, the magnetization of a crystal
of molecular clusters should relax 
at short times with a square-root time dependence
which is due to a gradual modification 
of the dipole fields in the sample caused by the tunneling
\begin{equation}
	M(H_z,t)=M_{\rm in}+(M_{\rm eq}(H_z)-M_{\rm in})
	       \sqrt{\Gamma_{\rm sqrt}(H_z)t} 
\label{M_sqrt}
\end{equation}
Here $M_{\rm in}$ is the initial magnetization at time $t$~= 0 
(after a rapid field change), 
and $M_{\rm eq}(H_z)$ is the equilibrium magnetization at $H_z$.
Experimentally, $M_{\rm eq}$ is difficult to measure and
we replaced it by the field cooled magnetization $M_{\rm FC}(H_z)$
(Fig.~\ref{hyst}a)
Intermolecular exchange interactions are neglected in the theory
of Prokof'ev and Stamp.

The rate function $\Gamma_{\rm sqrt}(H_z)$ is 
proportional to the normalized distribution $P(H_z)$ of molecules 
which are in resonance at $H_z$
\begin{equation}
	\Gamma_{\rm sqrt}(H_z) = 
	        c \frac{\xi_0}{E_D}
	          \frac{\Delta_{\pm S}^2}{4 \hbar} P(H_z) 
\label{gamma_sqrt}
\end{equation}
where $\xi_0$ is the line width coming 
from the nuclear spins, 
$E_D$ is the Gaussian half-width of $P(H_z)$, 
and $c$ is a constant of the order 
of unity which depends on the sample shape. 
Hence, the measurements of the 
short time relaxation as a 
function of the applied field $H_z$ give directly 
the distribution $P(H_z)$.

\begin{figure}
\begin{center}
\includegraphics[width=.45\textwidth]{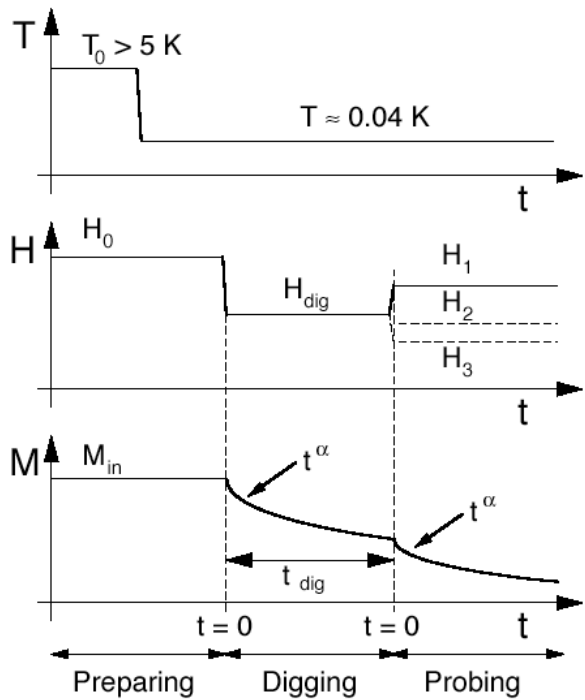}
\caption{Scheme of the hole digging method presenting
the time dependence of temperature, applied field, and magnetization of
the sample.}
\label{fig_dig}
\end{center}
\end{figure}

Motivated by the Prokof'ev--Stamp theory~\cite{Prokofev98},
we developed a new technique---which we called the
{\it hole digging method}---that can be 
used to observe the time evolution of
molecular states in crystals of nanomagnets~\cite{WW_PRL99}
and to establish resonant tunneling in systems
where quantum steps are smeared out by
small distributions of molecular environment~\cite{Soler04}.
Here, it has allowed us to measure the statistical 
distribution of magnetic bias
fields in the Mn$_4$ system that 
arise from the weak dipole and exchange fields of 
the clusters. A
hole can be ``dug'' into the distribution 
by depleting the available spins at
a given applied field.
Our method is based on the simple idea that after 
a rapid field change, the resulting short time relaxation 
of the magnetization is directly related to the
number of molecules which are in resonance at the given applied field.
Prokof'ev and Stamp have suggested that the short time relaxation should
follow a $\sqrt{t}-$relaxation law [equation~(\ref{M_sqrt})]. 
However, the hole digging method should
work with any short time relaxation
law---for example, a power law
\begin{equation}
	M(H_z,t)=M_{\rm in}+(M_{\rm eq}(H_z)-M_{\rm in})
	       (\Gamma_{\rm short}(H_z)t)^\alpha 
\label{M_power}
\end{equation}
where $\Gamma_{\rm short}$ is a characteristic short time
relaxation rate that is directly related to the
number of molecules which are in resonance at 
the applied field $H_z$, and $0 < \alpha <  1$ in most cases.
$\alpha$ = 0.5 in the Prokof'ev--Stamp theory [equation~(\ref{M_sqrt})]
and $\Gamma_{\rm sqrt}$ is directly proportional 
to $P(H_z)$ (Eq.~\ref{gamma_sqrt}).
The {\it hole digging method} can be divided into three steps 
(Fig.~\ref{fig_dig}):

\begin{enumerate}
\item
{\bf Preparing the initial state.} 
A well-defined initial magnetization 
state of the crystal of molecular 
clusters can be achieved by rapidly 
cooling the sample from high 
down to low temperatures in a constant applied 
field $H_z^0$. For zero applied 
field ($H_z$ = 0) or rather 
large applied fields ($H_z$~$>$~1~T), one 
yields the demagnetized or saturated 
magnetization state of the 
entire crystal, respectively. 
One can also quench the sample 
in a small field of a few milliteslas
yielding any possible initial magnetization $M_{\rm in}$. 
When the quench is fast 
($<<$~1~s), the sample's magnetization does 
not have time to relax, either by 
thermal or by quantum transitions. 
This procedure yields a frozen thermal 
equilibrium distribution,
whereas for slow cooling rates 
the molecule spin states in the crystal may tend to
a partially ordered ground state. 
Sect.~\ref{ordering} shows that, for our fastest
cooling rates of $\sim$ 1 s, partial ordering occurs.
However, we present a LZ-demagnetization method 
allowing us to reach a randomly disordered state.

\item
{\bf Modifying the initial state---hole digging.}  
After preparing the initial state, a field $H_{\rm dig}$ is applied 
during a time $t_{\rm dig}$, called {\it digging field} and 
{\it digging time}, 
respectively. During the digging time and depending on $H_{\rm dig}$, 
a fraction of the molecular spins tunnel (back and/or forth); that is, 
they reverse the direction of magnetization. 
\footnote{The field sweeping
rate to apply $H_{\rm dig}$ should be fast enough to minimize the change
of the initial state during the field sweep.}

\item
{\bf Probing the final state.} 
Finally, a field $H_z^{\rm probe}$ 
is applied (Fig.~\ref{fig_dig}) 
to measure the 
short time relaxation $\Gamma_{\rm short}$ 
(Eq.~\ref{M_power}) which is related to the number of spins 
that are still free for tunneling after step (2).
\end{enumerate}

The entire procedure is then repeated many times but at other 
fields $H_z^{\rm probe}$ yielding 
$\Gamma_{\rm short}(H_z,H_{\rm dig},t_{\rm dig})$
which is related to the distribution of spins 
$P(H_z,H_{\rm dig},t_{\rm dig})$
that are still free for tunneling after the hole digging.
For $t_{\rm dig}$ = 0, this method maps out the initial distribution.

We applied the hole digging method to several samples
of molecular clusters and quantum spin glasses. 
The most detailed study has been
done on the Fe$_8$ system. We found the 
predicted $\sqrt{t}$ relaxation (Eq.~(\ref{M_sqrt}) in 
experiments on fully saturated Fe$_8$ crystals \cite{Ohm98a,Ohm98b} 
and on nonsaturated samples \cite{WW_PRL99}. 
These results were in good agreement with 
simulations~\cite{Cuccoli99,Fernandez01,Tupitsyn04,Fernandez04,Tupitsyn05,Tupitsyn05}.

\begin{figure}
\includegraphics[width=.45\textwidth]{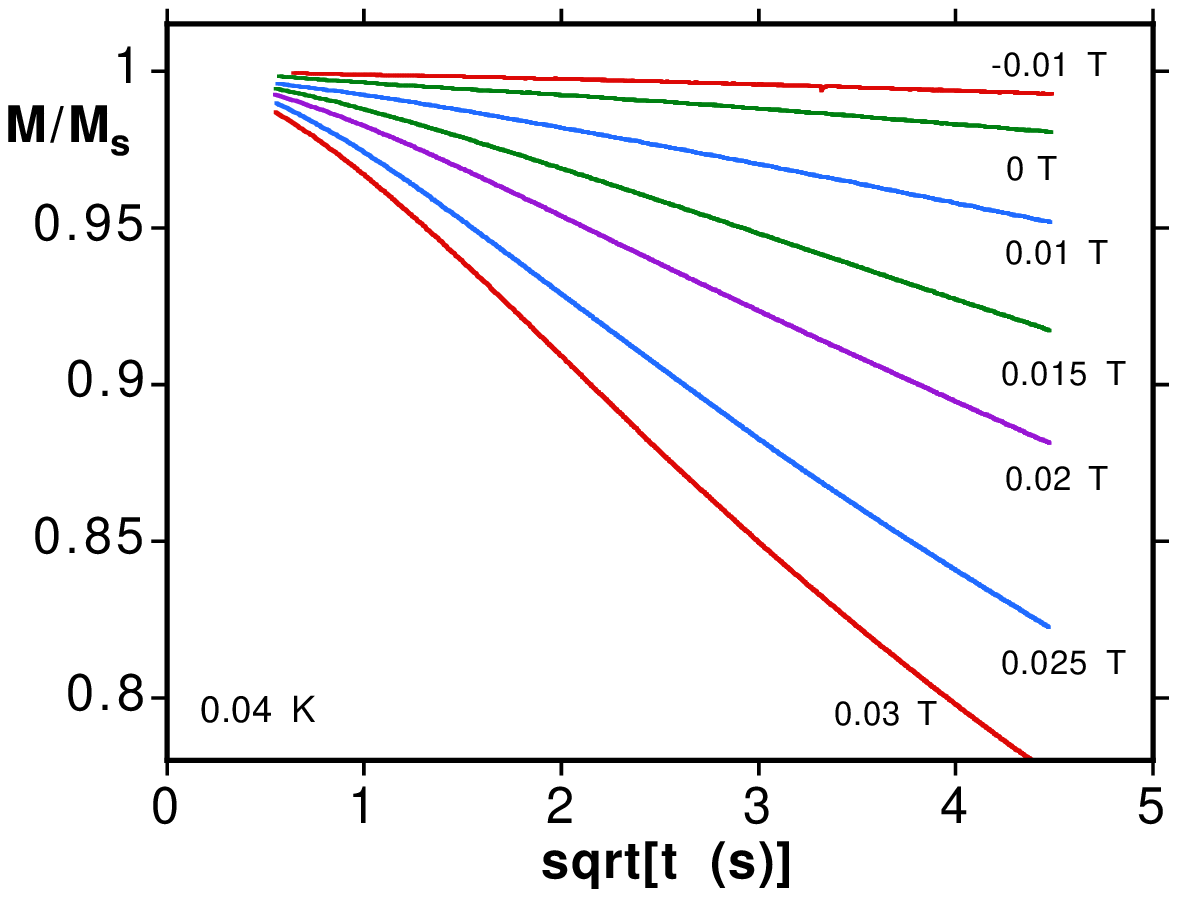}
\includegraphics[width=.45\textwidth]{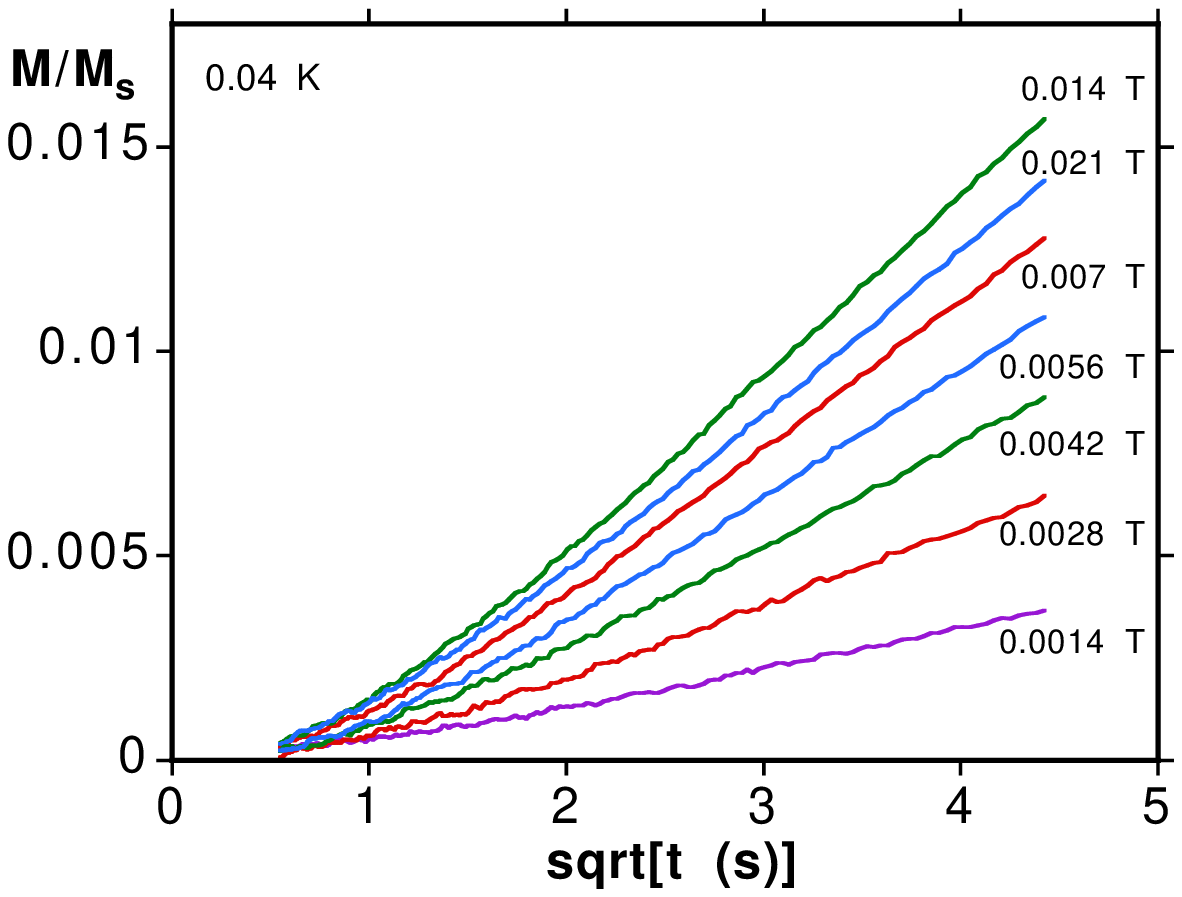}
\caption{(color online)  Typical square root of time relaxation curves for an Mn$_4$ 
crystal measured at 40~mK. For each curve, the sample was first 
(a) saturated or (b) thermally annealed at $H$~= 0. 
Then the indicated field was applied 
and the short time relaxation of magnetization was measured. The 
slope of the lines gives $\Gamma_{\rm sqrt}$ when plotted against the 
square-root of $t$ as shown.}
\label{M_t_sqrt}
\end{figure}

\subsection{Hole digging applied to Mn$_4$}
\label{digging_Mn4} 

\begin{figure}
\begin{center}
\includegraphics[width=.45\textwidth]{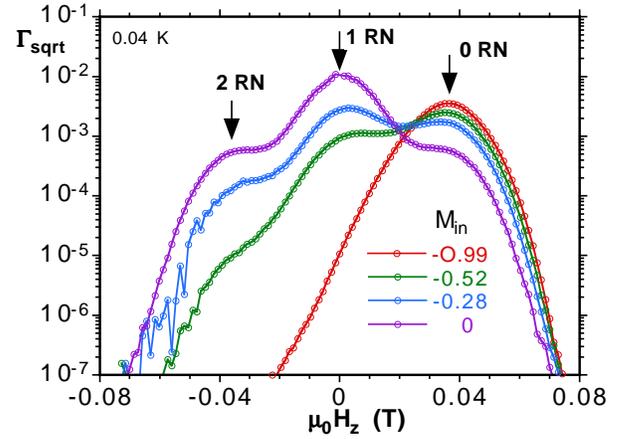}
\caption{(color online)  Field dependence of the short time square-root relaxation 
rates $\Gamma_{\rm sqrt}(H_z)$ for three different values of the 
initial magnetization $M_{\rm in}$. According to equation~(\ref{gamma_sqrt}), the 
curves are proportional to the distribution $P(H_z)$ of magnetic 
energy bias due to local internal field distributions in the 
sample. Note the logarithmic scale for $\Gamma_{\rm sqrt}$. The 
peaked distribution labeled $M_{\rm in}=-0.99 M_{\rm s}$ was obtained by 
saturating the sample, whereas the other distributions were obtained 
by thermal annealing. For $M_{\rm in} << M_{\rm s}$, the curves are distorted 
by nearest neighbor effects. 
The peak at $\pm0.036$ and 0 T are from molecules 
which have zero, one, or two nearest-neighbors (RN) molecules 
with reversed magnetization.}
\label{dist_M_in}
\end{center}
\end{figure}

Fig.~\ref{M_t_sqrt} shows typical relaxation curves plotted 
against the square-root of time. For initially saturated
or thermally annealed magnetization, the 
short time square root law is well obeyed. 
A fit of the data to Eq.~(\ref{M_sqrt}) determines $\Gamma_{\rm sqrt}$.
We took $M_{\rm eq} = M_{\rm FC}(H_z)$ of Fig.~\ref{hyst}a.
A plot of $\Gamma_{\rm sqrt}$ versus $H$ is shown in Fig.~\ref{dist_M_in} for 
the saturated samples ($M_{\rm in} \approx M_{\rm s}$), 
as well as for three other 
values of the initial magnetization which were obtained by quenching 
the sample from 5 K to 0.04 K in the presence of
a small field. The distribution 
for an initially saturated magnetization is clearly 
the most narrow reflecting the high degree of order starting from this 
state. The distributions become broader as the initial 
magnetization becomes smaller reflecting the random fraction of 
reversed spins. However, a clear fine structure emerges
with bumps at $\pm0.036$~T and  zero field which are
due to flipped nearest neighbor spins 
(Sect.~\ref{fine_structure_of_three}).

\begin{figure}
\includegraphics[width=.45\textwidth]{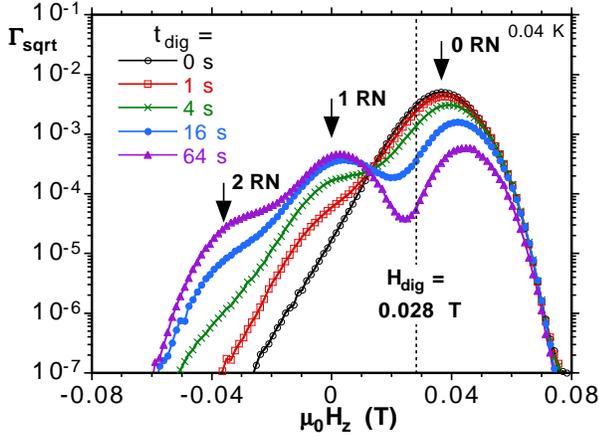}
\caption{(color online)  The field dependence of the short 
time square root relaxation rates $\Gamma_{\rm sqrt}(H_z)$ are 
presented on a logarithmic scale showing the depletion of the 
molecular spin states by quantum tunneling at $H_{\rm dig}$~= 0.028 T for various 
digging times $t_{\rm dig}$ and $M_{\rm in}=-M_s$.}
\label{dist_dig}
\end{figure}

Fig.~\ref{dist_dig}(a) shows the short time relaxation
rate for a digging field $H_{\rm dig}$~= 0.028 T 
and for several waiting times.
Note the rapid depletion of molecular spin 
states around $H_{\rm dig}$ and how quickly the 
same fine-structure, observed in Fig.~\ref{dist_M_in}, appears.
The hole  arises because only spins in resonance can tunnel. The hole is spread 
out because, as the sample relaxes, the internal fields in the sample 
change such that spins which were close to the resonance condition may 
actually be brought into resonance. The overall features are 
similar to experiments on a fully saturated Fe$_8$ crystals~\cite{WW_PRL99}.
However, the small chain-like intermolecular interactions
make the Mn$_4$ system unique for a deeper study presented in the 
following.

\subsubsection{Chain-like intermolecular interactions}
\label{fine_structure_of_three}
The Mn$_4$ molecules are arranged along the $c-$axis in a chain-like 
structure (Fig.~\ref{structure_cell_SB1}). The dipolar coupling between
molecules along the chain is significantly larger than
between molecules in different chains. In addition there
are small exchange coupling between molecules along
the chain (Sect.~\ref{structure}), leading us to propose
the following model. 

\begin{figure}
\includegraphics[width=.45\textwidth]{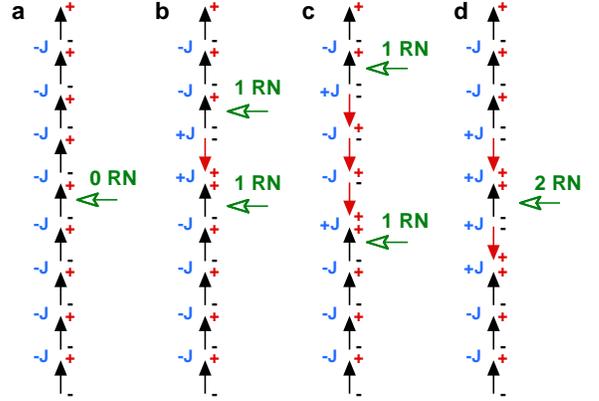}
\caption{(color online)  Schematical representation of a chain
of spins with dipole and exchange coupling represented
by $+$ and $-$ poles and $\pm J$, respectively.
(a) All spins have zero reversed neighbors (0 RN);
in (b) and (c), the indicated spins have one
reversed neighbor (1 RN), whereas in (d) it
has two reversed neighbors (2 RN).
The three cases of this two-neighbor-model
lead to the fine structure of three for all
quantum resonance steps.}
\label{schema_RN}
\end{figure}

Each arrow in Fig.~\ref{schema_RN}
represents a molecule. The $+$ and $-$ signs are
the magnetic poles. The exchange coupling is represented
by $\pm J$. The ground state for a ferromagnetic
chain is when all spins are up or down with $+$
and $-$ poles together and $-J$ for all exchange
couplings (Fig.~\ref{schema_RN}a). 
For short, we say that all spins have zero reversed neighbors (2 RN).
In order to reverse 
one spin at its zero-field resonance ($m = -S$ and $m' = S$), 
a magnetic field
has to be applied that compensates the interaction 
field from the neighbors. 
As soon as one spin is reversed  
(Fig.~\ref{schema_RN}b),
the two neighboring spins see a positive interaction field
from one neighbor and a negative one from the other neighbor,
that is we say for short that 
the two spins have one reversed neighbor (1 RN).
The interaction field seen by those spins is compensated.
Such a spin with 1 RN has a resonance at zero applied field
and might reverse creating another spin with 1 RN 
(Fig.~\ref{schema_RN}c). The third case is when a spin
has 2 RN (Fig.~\ref{schema_RN}d). In this case,
a negative field has to be applied to compensate the
interaction field of the two neighbors.

In summary, there are three possibilities for a given spin:
0 RN, 1RN, or 2RN with a zero-field resonance shifted to
positive (0 RN), zero (1 RN), or negative (2 RN) fields.
The influence of the interaction fields of the
neighboring molecules is taken into account by a bias field
$H_z^{\rm bias}$. The effective field $H_{z}$ acting on the molecule is
therefore the sum of the applied field $H_z^{\rm app}$ and the bias field
$H_z^{\rm bias}$:
\begin{equation}
	H_{z} = H_{z}^{\rm app} + H_{z}^{\rm bias} = H_{z}^{\rm
	app} + \frac {1}{g\mu_{B}\mu_{0}}
	\sum_{k=1}^{2}J_{\rm eff}M_{k}
\label{eq_H_bias}
\end{equation}
where $M_k$ is the quantum number of the neighboring molecule 
and $J_{\rm eff}$ is an effective exchange coupling taking
into account of the nearest neighbor exchange and dipolar coupling.
$J_{\rm eff} \approx$~0.01~K for Mn$_4$.

\begin{figure}
\includegraphics[width=.45\textwidth]{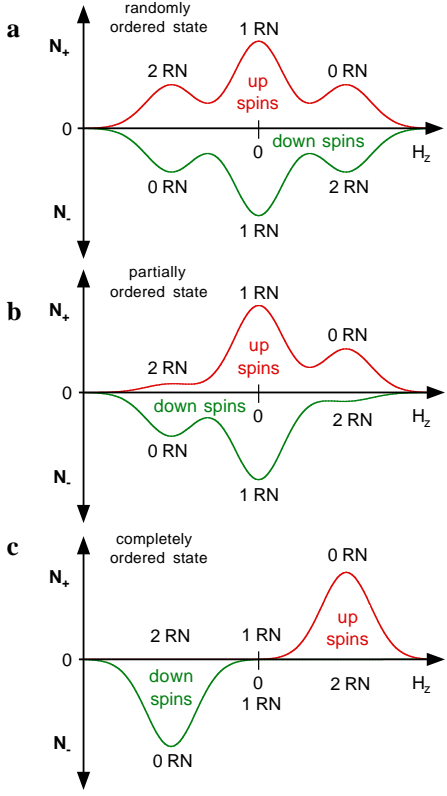}
\caption{(color online)  Schema of the distributions
of internal fields of (a) a randomly ordered, (b) a partially ordered,
and (c) a completely ordered state with zero total magnetization.
Here, $N_+$ and $N_-$ are the distributions for up and down spins,
respectively. The fine structure with three bumps are due to
three cases of zero reversed neighbors (0 RN), one
reversed neighbor (1 RN), and two reversed neighbors (2 RN).
}
\label{schema_dist}
\end{figure}

Because of the long range character of dipolar fields and
the interchain dipolar couplings, the situation in 
a Mn$_4$ crystal is more complicated. However, when the 
exchange interaction is significantly larger than the
dipolar interaction, the long range character of the latter
leads only to a broadening of the two-neighbor model.
Fig.~\ref{schema_dist} presents schematically the distribution
of internal fields of a randomly ordered, a partially ordered,
and a completely ordered state with zero total magnetization.
Such distributions can be observed with a short-time
relaxation, presented in Figs.~\ref{dist_M_in} 
for a Mn$_4$ crystal with different initial magnetizations.

We tested the two-neighbor model
extensively using, for example, minor hysteresis loops and
starting from an initially saturated state (Fig.~\ref{schema_RN}a).
When sweeping the field over the zero-field 
resonance after a negative saturation field,
resonant tunneling can only occur at the positive
interaction field of 0.036 T. The corresponding step in $M(H)$
indicates that few spins with 0 RN reversed creating spins
with 1 RN. This leads to two steps when sweeping 
the field backwards over the zero-field resonance, one for 0 RN and
one for 1 RN (Fig.~\ref{hyst_min}b). When enough spins
are reversed, a third step appears at 2 RN.

Similar experiments can be done with the hole digging
method (Sect.~\ref{digging}). Digging a hole at the field of 0 RN induces
a peak at 1 RN (Fig.~\ref{dist_dig}).

\subsubsection{Magnetic ordering in crystals of single-molecule magnets}
\label{ordering} 

The question of magnetic ordering in molecular magnets has recently been 
addressed theoretically~\cite{Chudnovsky01b,Fernandez02}.
Depending on the system ferromagnetic, antiferromagnetic, or
spin glass like ground states with ordering temperatures
between about 0.2 and 0.5 K have been predicted.
Due to the slow relaxation of SMMs at low temperature,
ordering might happen at non-accessible long time scales.
Recent experimental studies concerned 
antiferromagnetic ordering in Fe$_{19}$ SMMs~\cite{Affronte02}, 
ferromagnetic ordering of high-spin molecules~\cite{Morello03},
and partial ordering in the fast tunneling regime of SMMs~\cite{Evangelisti04}.
We present here a simple method to show that partial ordering
occurs in crystals of Mn$_4$ SMMs in the slowly tunneling regime.

The first important step is to create a randomly disordered 
state (Fig.~\ref{schema_dist}a), 
that is for any internal field value there are the same number of
up and down spins. This means that for any applied field,
no magnetization relaxation can be observed because the
tunneling from up to down is compensated by tunneling from
down to up.

\begin{figure}
\includegraphics[width=.45\textwidth]{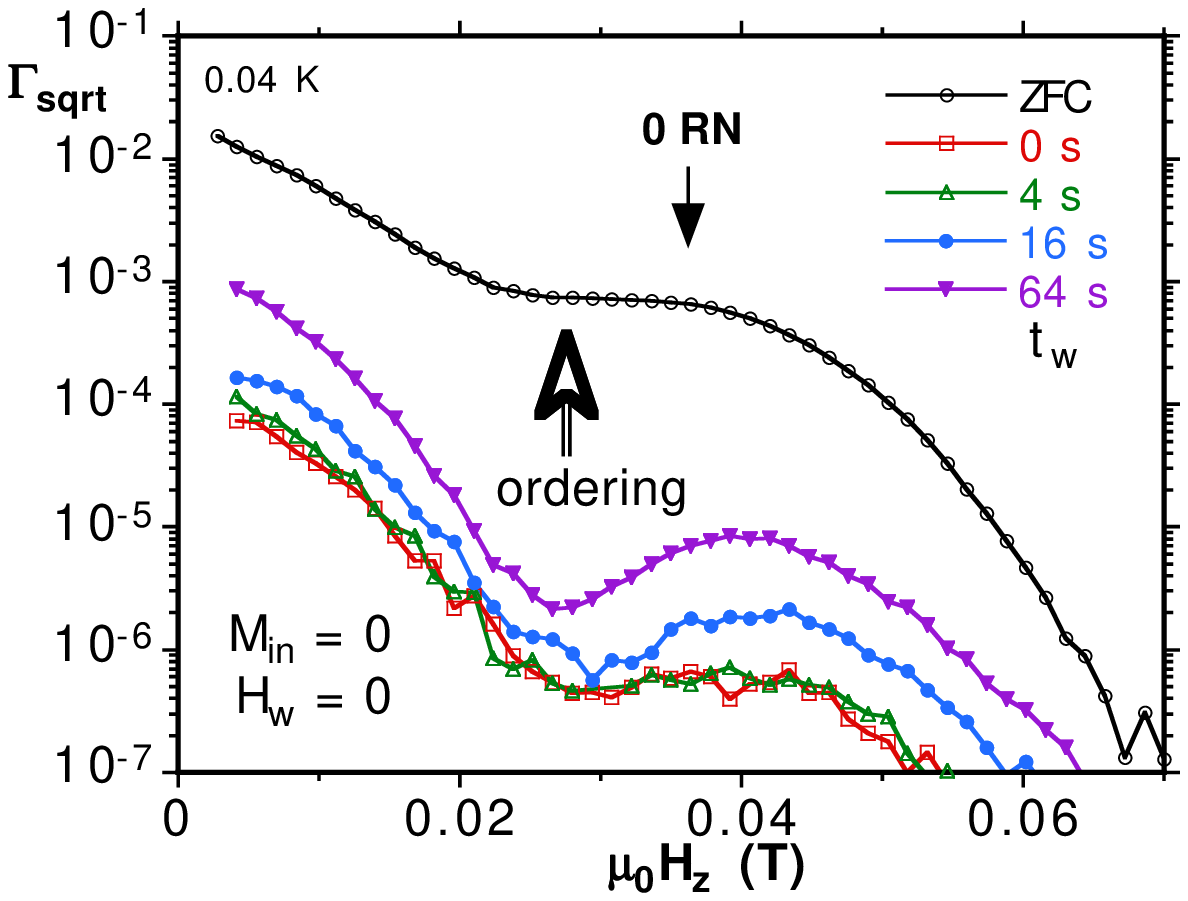}
\includegraphics[width=.45\textwidth]{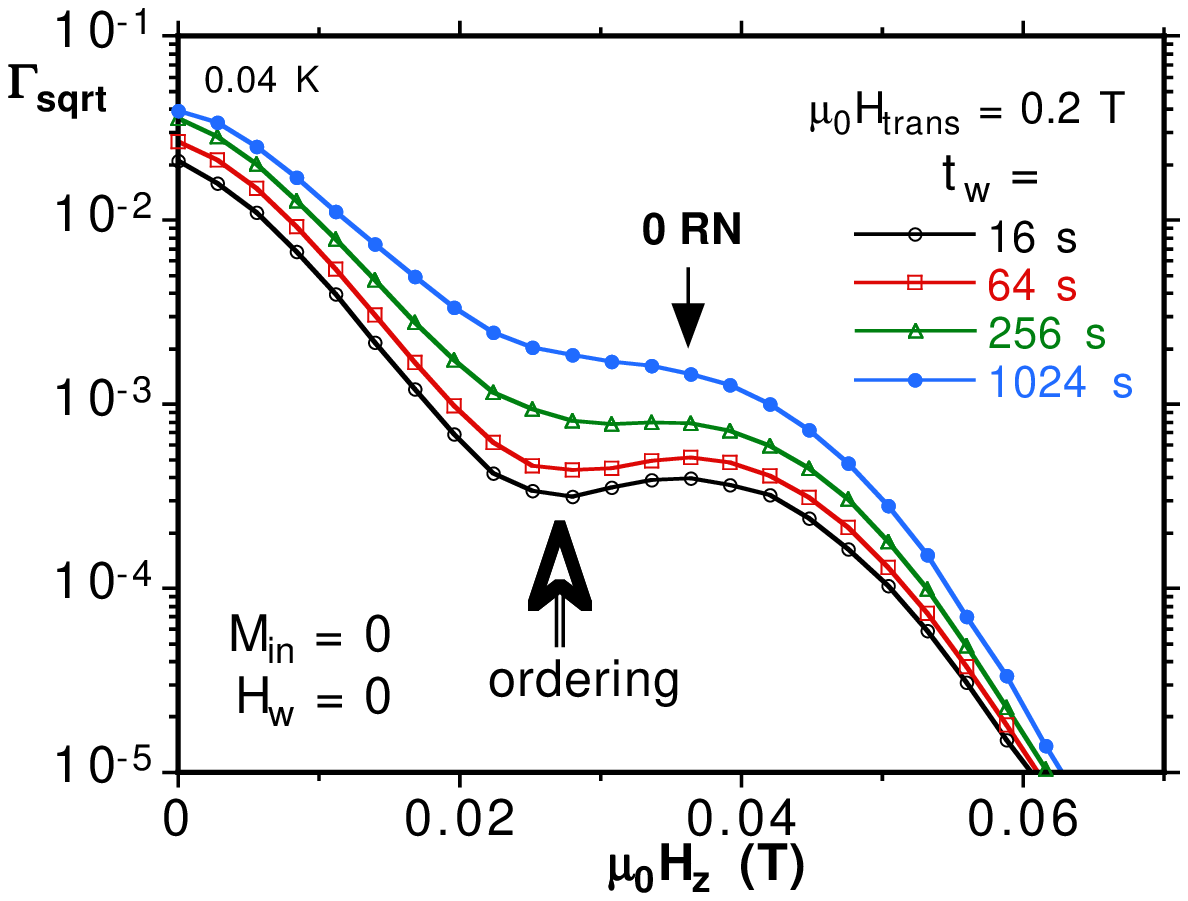}
\caption{(color online)  Short time square root relaxation rate for an Mn$_4$ 
crystal measured at 40~mK. (a) The ZFC curve was measured 
after thermal annealing at $H$~= 0.
For the other curves, a randomly disordered 
state with $M_{\rm in} = 0$ was first created
by sweeping back and forth the field over
the zero-field resonance. After a waiting
time $t_{\rm w}$ at $H$~= 0, the short time 
relaxation rate was measured.
The field for one reversed neighbor (1 RN) is indicated.
For longer $t_{\rm w}$, $\Gamma_{\rm sqrt}$ approaches
the ZFC curve of a partially ordered state.
(b) Similar to Fig.~\ref{dist_ordering}a but during the waiting time 
a transverse field of 0.2 T was applied leading to
faster ordering.}
\label{dist_ordering}
\end{figure}

We tried to achieve a randomly disordered 
state (Fig.~\ref{schema_dist}a) by a fast
quench of the sample temperature from 5 K
down to 0.04 K ($\sim$~1~s). When applying a small
field, a $\sqrt{t}$-relaxation is observed (Fig.~\ref{M_t_sqrt}b)
showing the the sample was already partially
ordered (Fig.~\ref{schema_dist}b).

We found that a randomly disordered 
state (Fig.~\ref{schema_dist}a) can be achieved
by sweeping back and forth the field over
the zero-field resonance. During each sweep,
few spins tunnel randomly back and fourth. When the Landau--Zener
tunnel probability is small ($P_{\rm LZ} << 1$), that is
for fast sweep rates of 0.1 T/s for Mn$_4$, and a large
number of back and forth sweeps, a magnetization
state can be prepared that shows only a very small
relaxation when applying a small field (Fig.~\ref{dist_ordering}a).
Ordering can then be observed by simply waiting at
$H$ = 0 for a waiting time $t_{\rm w}$.
The longer is $t_{\rm w}$, the larger is the relaxation,
that is the distribution of internal fields evolves
from a randomly disordered 
state (Fig.~\ref{schema_dist}a) to a partially ordered 
state (Fig.~\ref{schema_dist}b).
In order to enhance the ordering, a transverse field
can be applied during the waiting time (Fig.~\ref{dist_ordering}a).
Note that we did not observe complete ordering (Fig.~\ref{schema_dist}c) 
which is probably due to the entropy, similar to an infinite spin 
chain which will not order at $T=0$ due to entropy. 
It is also interesting to note that ordering does not 
quench tunneling.

\begin{figure}
\begin{center}
\includegraphics[width=.45\textwidth]{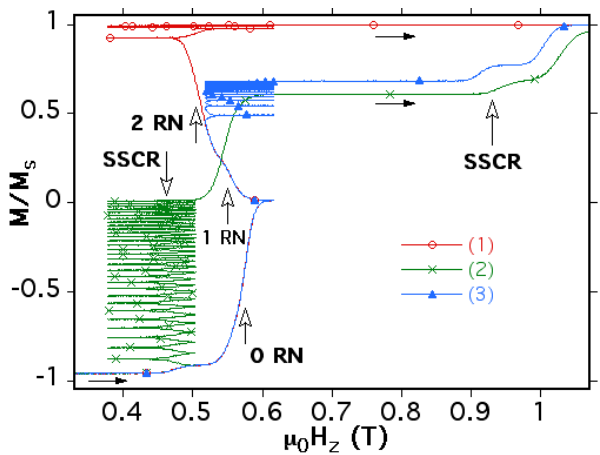}
\caption{(color online)  Minor hysteresis loops of a single crystal of Mn$_4$.
The magnetization was first saturated at -1.4 T. 
After ramping the field to zero, the field
was swept several times back and forth 
over a part of the tunneling step at the 
level crossing $(m,m')=(-9/2,7/2)$:
for curve (1) between 0.38 and 0.62 T;
for curve (2) between 0.38 and 0.5 T;
for curve (3) between 0.52 and 0.62 T.
After the back and forth sweeps, the field is
swept to 1.4 T. The field sweep rate for all
parts is 0.07 T/s.
The fine structure due to nearest neighbor effects
and SSCR are indicated.}
\label{hyst_min_SSCR}
\end{center}
\end{figure}

\section{Spin-spin cross-relaxation in single-molecule magnets}
\label{SSCR} 

We showed recently that the one-body tunnel 
picture of SMMs (Sect.~\ref{LZ}) is not always sufficient 
to explain the measured tunnel transitions. An improvement 
to the picture was proposed 
by including also two-body tunnel transitions 
such as spin-spin cross-relaxation (SSCR) which are mediated
by dipolar and weak exchange interactions between 
molecules~\cite{WW_PRL02}. 
At certain external fields, SSCRs lead to additional quantum
resonances which show up in hysteresis loop measurements 
as well defined steps. A simple model was used to explain 
quantitatively all observed transitions.
\footnote{We used different techniques to show that 
different species due to loss of solvent or
other defects are not the reason of the observed additional 
resonance transitions.}
Similar SSCR processes were also 
observed in the thermally activated regime of a LiYF$_4$ 
single crystal doped with Ho ions~\cite{Giraud01}
and for lanthanide SMMs~\cite{Ishikawa05a}.

In order to obtain an approximate understanding 
of SSCR, we considered 
a Hamiltonian describing two coupled SMMs
which allowed as to explain quantitatively
13 tunnel transitions.
We checked also that all 13 transitions
are sensitive to an applied transverse field, which
always increases the tunnel rate. 
The parity of the level crossings was also established 
and in agreement with the two-spin model~\cite{WW_PRL02}.

It is important to note that 
in reality a SMM is coupled to many 
other SMMs which in turn are coupled to many other 
SMMs. This represents a complicated many-body problem
leading to quantum processes involving more than two SMMs. 
However, the more SMMs that are involved, the lower 
is the probability for occurrence. 
In the limit of small exchange couplings and transverse
terms, we therefore consider only processes 
involving one or two SMMs. 
The mutual couplings between all SMMs 
should lead mainly to broadenings and small 
shifts of the observed quantum steps
which can be studied with minor hysteresis loops.

Fig.~\ref{hyst_min_SSCR} shows typical minor loops at 
the level crossing $(m,m')=(-9/2,7/2)$.
In curve (1), the field is swept forth and back
over the entire
resonance transition. After about two forth and back
sweeps, all spins are reversed. Note the
next-nearest neighbor fine structure that is
in prefect agreement with the two-neighbor model
(Sect.~\ref{fine_structure_of_three}).
In curve (2), the field is swept forth and back
over a SSCR transition (transition 7 in reference
~\cite{WW_PRL02}). Note that the relaxation rate is
much slower because of the low probability
of SSCRs and the fact that this transition is mainly
possible for spins with 0 RN or 1 RN.
In curve (3), the field is swept forth and back
over a part of the level crossing $(m,m')=(-9/2,7/2)$
corresponding to spins with 0 RN or 1 RN.
In this case, the relaxation rate decreases
strongly after the first forth and back sweep
because the 2 RN spins cannot tunnel in this
field interval.

\section{Conclusion}
Resonance tunneling measurements on a new 
high symmetry Mn$_{4}$ molecular nanomagnet show 
levels of detail not yet possible with other
SMMs, as a result of higher symmetry and
a well isolated spin ground state of $S = 9/2$. 
This has permitted an 
unprecedented level of analysis of the data to be 
accomplished, resulting in information not yet attainable 
with other SMMS. 
In particular, Landau--Zener (LZ) tunneling 
in the presence of weak intermolecular dipolar and 
exchange interactions can be studied, 
using the LZ and inverse LZ method. The latter
has not been applied to any other SMM.
Three regions are identified: (i) at small transverse fields,
tunneling is dominated by single tunnel transitions;
(ii) at intermediate transverse fields, the
measured tunnel rates are governed by reshuffling of internal fields,
(iii) at larger transverse fields, the magnetization reversal
starts to be influenced by the direct relaxation process
and many-body tunnel events might occur.
The hole digging method is used to study the next-nearest neighbor 
interactions.
At small external fields, it is shown that magnetic ordering occurs which
does not quench tunneling. An applied transverse field can 
increase the ordering rate. Spin-spin cross-relaxations, mediated
by dipolar and weak exchange interactions, are proposed to
explain additional quantum steps.
We would like to emphasize that
the present study is mainly experimental, aiming to
encourage theorists to develop new tools to model 
the quantum behavior of weakly
interacting quantum spin systems.




\end{document}